\documentclass{article}

\usepackage{arxiv}

\usepackage[utf8]{inputenc} 
\usepackage[T1]{fontenc}    
\usepackage{hyperref}       
\usepackage{url}            
\usepackage{booktabs,longtable}       
\usepackage{siunitx}
\usepackage{amsfonts}       
\usepackage{nicefrac}       
\usepackage{microtype}      
\usepackage{lipsum}
\usepackage{xcolor}
\usepackage{float}

\providecommand{\tightlist}{%
	\setlength{\itemsep}{0pt}\setlength{\parskip}{0pt}}
\usepackage{mathtools}
\usepackage{float}
\usepackage[style=nature,backend=biber,doi=false,url=false]{biblatex}
\title{Seeding Method for Ice Nucleation under Shear}

\author{
	\href{https://orcid.org/0000-0001-8706-2383}{\includegraphics[scale=0.06]{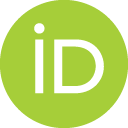}\hspace{1mm}Amrita Goswami}\\
  Department of Chemical Engineering\\
  Indian Institute of Technology Kanpur\\
  \texttt{amritag@iitk.ac.in} \\
	\And
	{Indranil Saha Dalal}\thanks{\textbf{Corresponding Author}} \\
  Department of Chemical Engineering\\
  Indian Institute of Technology Kanpur\\
  \texttt{indrasd@iitk.ac.in} \\
	\And
	\href{https://orcid.org/0000-0001-8056-2115}{\includegraphics[scale=0.06]{orcid.png}\hspace{1mm}Jayant K. Singh}\footnotemark[1] \\
	Department of Chemical Engineering\\
	Indian Institute of Technology Kanpur\\
	\texttt{jayantks@iitk.ac.in} \\
}

\begin{document}
\maketitle

\begin{abstract}
	Hydrodynamic flow can have complex and far-reaching consequences on the
  rate of homogenous nucleation. We present a general formalism for
  calculating the nucleation rates of simply sheared systems. We have
  derived an extension to the conventional Classical Nucleation Theory,
  explicitly embodying the shear rate. Seeded Molecular Dynamics
  simulations form the backbone of our approach. The framework can be used
  for moderate supercoolings, at which temperatures brute-force methods
  are practically infeasible. The competing energetic and kinetic effects
  of shear arise naturally from the equations. We show how the theory can
  be used to identify shear regimes of ice nucleation behaviour for the mW
  water model, unifying disparate trends reported in the literature. At
  each temperature, we define a crossover shear rate in the limit of
  \(1000-10,000 \ s^{-1}\), beyond which the nucleation rate increases
  steadily upto a maximum, at the optimal shear rate. For \(235\),
  \(240\), \(255\) and \(260 \ K\), the optimal shear rates are in the
  range of \(\approx 10^6-10^7 \ s^{-1}\). For very high shear rates
  beyond \(10^8 \ s^{-1}\), nucleation is strongly inhibited. Our results
  indicate that the shear-dependent nucleation rate curves have a
  non-monotonic dependence on temperature.
\end{abstract}


\hypertarget{introduction}{%
\section{Introduction}\label{introduction}}

Crystal formation under the action of shear flow is a ubiquitous
phenomenon, with important implications in nature and various medical,
metallurgical and industrial applications
\cite{Penkova2006, Woodhouse2013, Berland1992, Baird2011, Forsyth2014, KeltonGreer2010}.
Hydrodynamic flows can induce complex changes in the behaviour of
nucleating systems, ranging from shear-mediated ordering of the liquid
\cite{Ackerson1988, Yan1994, Haw1998, Amos2000} to disruption of
crystallization \cite{Palberg1995, Okubo1999}.

The literature is rife with contradictory accounts of the effects of
shear on the nucleation rate. Certain experiments surmise that shear
flow can retard nucleation \cite{Blaak2004, Blaak2004a}, while others
report that shear flows have a negligible influence on crystallization
\cite{Hansen2001, Akio1992, 2017a}. Different studies have shown that
shear flow can enhance the rate of nucleation
\cite{Cerda2008, Mokshin2009, Graham2009, Shao2015, Ruiz-Franco2018}. On
the other hand, simulation studies suggest that the nucleation rates
have a non-monotonic dependence on the applied shear rates, and exhibit
a maximum at an optimal shear rate
\cite{Mokshin2013, Richard2015, Mura2016, Luo2020}.

Even without the added variable of shear, the use of computer
simulations for describing homogenous crystal nucleation is complicated
by the fact that it is a rare event, occurring on time scales
inaccessible to conventional Molecular Dynamics (MD). Techniques of
analyzing crystal nucleation, dependent on brute-force simulations, are
typically only feasible for deeply supercooled systems. This is partly
due to the stochastic nature of nucleation events. In practice,
obtaining sufficiently good statistics from brute-force MD-based methods
can be intractable for moderate supercooling \cite{Sosso2016a}.

The event at the crux of crystal nucleation, which is an activated
process, is the formation of a sufficiently large crystalline cluster in
the metastable liquid phase to overcome the free-energy barrier. Seeded
MD simulations can be used to reliably estimate the critical nucleus
size and the interfacial energy
\cite{Espinosa2014, Espinosa2015, Espinosa2016}. In the quintessential
seeding method, the information gleaned from seeded simulations is used
in conjunction with Classical Nucleation Theory (CNT)
\cite{Volmer1926, Becker1935} to calculate the free-energy barrier and
nucleation rate. Thus, the seeding technique can be particularly
advantageous for moderate supercoolings, at which temperatures the
spatiotemporal resolution of brute-force MD can be insufficient to
observe nucleation events.

In this work, we extend CNT to incorporate shear, and propose a general,
computationally efficient framework that builds on the seeding method.
We study the effect of a wide range of shear rates on the ice nucleation
rates at different temperatures, for which brute-force MD and
equilibrium rare-event sampling methods are essentially infeasible. We
show that the formalism can pinpoint distinct regimes of ice nucleation
behaviour, and is thus able to consolidate and justify the seemingly
inconsistent results reported in the literature. The results suggest a
previously unexplored non-linear temperature dependence of the
nucleation rates.

\hypertarget{theory-and-methods}{%
\section{Theory and Methods}\label{theory-and-methods}}

\hypertarget{overview-of-the-seeding-method}{%
\subsection{Overview of the Seeding
Method}\label{overview-of-the-seeding-method}}

According to Classical Nucleation Theory (CNT)
\cite{Volmer1926, Becker1935, Sunyaev1992, Sosso2016a}, the free energy
of formation \(F\) of a spherical crystallite nucleus of radius \(R\)
can be expressed as the sum of a favourable volume term and an
unfavourable surface term:

\begin{equation}\label{eqn:cnt1}
F(R) = -\frac{4}{3} \pi R^3 \frac{|\Delta \mu|}{v'} + 4 \pi R^2 \sigma, \tag{$1$}
\end{equation}

where \(|\Delta \mu|\) is the chemical potential difference between the
metastable liquid phase and the crystal, \(v'\) is the volume of one
particle or molecule in the crystal, and \(\sigma\) is the crystal-fluid
interfacial free energy.

In the seeding method \cite{Bai2005, Bai2006a}, a solid cluster of the
crystal phase is inserted into the supercooled fluid, and the
temperature at which the cluster has a critical number of molecules
(\(N^*\)) is determined. For smaller clusters, the free energy cost of
forming the solid-liquid interface dominates and the embryos tend to
shrink. For clusters larger than the critical nucleus size, the volume
contribution is greater, favouring crystal growth.

The critical cluster size \(N^*\) is the cluster size which maximizes
the free energy in Eq.\eqref{eqn:cnt1}, given by

\begin{equation}\label{eqn:cntN}
N^* = \frac{32 \pi v'^2 \sigma^3}{3 |\Delta \mu |^3}, \tag{$2$}
\end{equation}

where \(N^*\) is the number of particles or molecules in the critical
cluster. The critical cluster corresponds to a particular temperature,
at which roughly half of the trajectories show crystal growth. This
implies that, for a known inserted cluster size, the temperature at
which the cluster is critical can be estimated.

Thus, the seeding technique directly yields \(N^*\) at the estimated
temperature \(T\). Once \(T\) has been fixed, \(|\Delta \mu|\) and
\(v'\) become fixed as well. The interfacial energy \(\sigma\), can now
be determined from Eq. \eqref{eqn:cntN} as:

\begin{equation}\label{eqn:cntInterfacial}
\sigma = \left( \frac{3 N^*}{32 \pi v'^2} \right)^{\frac{1}{3}} |\Delta \mu|. \tag{$3$}
\end{equation}

Putting Eq. \eqref{eqn:cntN} in Eq. \eqref{eqn:cnt1}, we obtain the free
energy barrier for nucleation:

\begin{equation}\label{eqn:cntFreeEnerg}
F(N^*) = \frac{16 \pi v'^2 \sigma^3}{3 |\Delta \mu |^2}, \tag{$4$}
\end{equation}

where \(F(N^*)\) is the height of the free energy barrier.

The rate of nucleation can be estimated as
\cite{Kelton1983, Kelton1991, Auer2001, Auer2004}

\begin{equation}\label{eqn:rateCNT}
J = \rho_l Z f^+ e^{-\frac{F(N^*)}{k_B T}}, \tag{$5$}
\end{equation}

where \((\rho_lZ f^+)\) is the kinetic pre-factor, in which \(\rho_l\)
is the number density of the supercooled liquid, \(Z\) is the Zeldovich
factor \cite{Kelton1991}, and \(f^+\) is the rate of attachment of
molecules to the cluster in units of inverse time. \(k_B\) is the
Boltzmann constant, and \(T\) is the temperature at which the inserted
seed of size \(N^*\) is critical.

The Zeldovich factor \(Z\) captures the multiple possible re-crossings
of the free energy barrier \cite{Pan2004} and is related to the
curvature of the free energy curve at the critical cluster size \(N^*\).
According to CNT, which assumes a perfectly spherical cluster, the
Zeldovich factor can be approximated by:

\begin{equation}\label{eqn:zeldovichCNT1}
Z = \sqrt{\frac{|F''(N^*)|}{2 \pi k_B T}} = \sqrt{\frac{|\Delta \mu|}{6\pi k_B T N^*}}, \tag{$6$}
\end{equation}

where \(F''(N^*) = \frac{\partial^2 F(N)}{\partial N^2}\Big|_{N=N^*}\).

Thus, \(Z\) can be directly calculated using
Eq.\eqref{eqn:zeldovichCNT1} from the quantities obtained from seeding
computations and the chemical potential difference between the liquid
and solid phases \(|\Delta \mu|\).

The main advantage of the seeding technique is that it allows for a
direct estimate of the interfacial free energy for a wide range of
supercooling conditions. Using the framework of CNT, the free energy
barrier and the nucleation rate can be calculated from quantities
obtained from seeded simulations, the chemical potential difference and
the crystal density.

In the subsequent sections, we extend the CNT equations for bulk
homogenous systems to account for volume-preserving shear, building on
the seeding method.

\hypertarget{extension-of-the-seeding-technique-and-cnt-for-sheared-systems}{%
\subsection{Extension of the Seeding Technique and CNT for Sheared
Systems}\label{extension-of-the-seeding-technique-and-cnt-for-sheared-systems}}

According to Mura and Zaccone \cite{Mura2016}, the free energy of a crystal nucleus in a
bulk nucleating system, subjected to a volume-preserving shear
(``simple'' shear) \(\dot{\gamma}\), is given by

\begin{equation}\label{eqn:freeEnergyR}
F(R) = -\frac{4}{3} \pi R^3 \frac{|\Delta \mu_{0} |}{v'} + 4 \pi R^2 \sigma_{0} \left( 1 + \frac{7}{24} \frac{\eta^2 \dot{\gamma}^2 }{G^2}\right) + \frac{1}{2} \frac{\eta^2 \dot{\gamma}^2 }{G} \frac{4}{3} \pi R^3, \tag{$7$}
\end{equation}

where \(F(R)\) is the free energy of formation of a cluster of radius
\(R\), \(|\Delta \mu_{0}|\) is the chemical potential difference between
the thermodynamically stable crystal phase and the metastable liquid
phase when no shear is applied, \(\sigma_{0}\) is the surface tension or
the interfacial free energy of the nucleus at zero shear, \(v'\) is the
volume of one molecule in the crystal phase, \(\eta\) is the fluid
viscosity, and \(G\) is the shear modulus of the nucleus.

Here, the term \(\frac{7}{24} \frac{\eta^2 \dot{\gamma}^2}{G^2}\) in
Eq.\eqref{eqn:freeEnergyR} is a ``shape factor'' correction; which
accounts for the deformation of the nucleus into an ellipsoid.

As noted in the previous section, the eponymous seeding technique
provides a direct estimate for the interfacial free energy
\(\sigma_{0}\), at the temperature \(T\) for which the inserted seed is
critical. Thus, it is desirable to convert the free energy in
Eq.\eqref{eqn:freeEnergyR} to a function of the cluster size \(N\).
Changing the variable \(R\) to \(N\) enables us to rewrite the free
energy as

\begin{equation}\label{eqn:freeEnergyN}
F(N) = -N \left( \frac{2G |\Delta \mu_{0}| - \eta^2 \dot{\gamma}^2 v'}{2G} \right) + \frac{\pi \sigma_{0}}{6G^2} \left( \frac{3v'N}{4\pi} \right)^{\frac{2}{3}} (24G^2 + 7\eta^2 \dot{\gamma}^2). \tag{$8$}
\end{equation}

The equation above has a stationary point corresponding to the critical
nucleus size \(N^*\). Solving for the stationary point of
Eq.\eqref{eqn:freeEnergyN}, we obtain:

\begin{equation}\label{eqn:criticalN}
N^* = \frac{\pi \sigma_{0}^3 v'^2}{162G^3} \left( \frac{24G^2 + 7\eta^2 \dot{\gamma}^2}{2G |\Delta \mu_{0}| - \eta^2 \dot{\gamma}^2 v'} \right)^3. \tag{$9$}
\end{equation}

The critical nucleus size corresponds to the stationary point of the
free energy surface, defined by Eq.\eqref{eqn:freeEnergyN}. Thus the
height of the free energy barrier for nucleation, corresponding to the
critical nucleus size \(N^*\), is given by

\begin{equation}\label{eqn:criticalF}
F(N^*) = \frac{\pi \sigma_{0}^3 v'^2}{648G^4} \frac{(24G^2+7\eta^2 \dot{\gamma}^2)^3}{(2G |\Delta \mu_{0}| - \eta^2 \dot{\gamma}^2v')^2}. \tag{$10$}
\end{equation}

We note that the free energy barrier increases with an increase in the
shear rate as governed by Eq.\eqref{eqn:criticalF}. By construction,
Eq.\eqref{eqn:criticalN} also predicts a concomitant increase in
\(N^*\).

\hypertarget{calculation-of-the-nucleation-rate}{%
\subsection{Calculation of the Nucleation
Rate}\label{calculation-of-the-nucleation-rate}}

The dynamics of a nucleation process can be formulated as a
Kramers-Moyal expansion \cite{Moyal1949, Kampen1992}, which can be
recast as a quintessential Focker-Planck equation \cite{Reguera2005}

\begin{equation}\label{eqn:fockerPlanck}
\frac{\partial P(N,t)}{\partial t} = \frac{\partial}{\partial N} \left( D e^{-\frac{F(N)}{k_B T}} \frac{\partial}{\partial N} \Big( P(N,t) \Big) e^{\frac{F(N)}{k_B T}} \right),  \tag{$11$}
\end{equation}

where \(P(N,t)\) is the probability distribution of the clusters
containing \(N\) molecules at time \(t\), and \(D\) is a diffusion
coefficient, or in this case, the rate of attachment of molecules to a
nucleus in units of inverse time. \(F(N)\) (Eq. \eqref{eqn:freeEnergyN})
is the free energy barrier for this nucleation process.

We recognize that Eq.\eqref{eqn:fockerPlanck} can be reformulated as the
Zeldovich-Frenkel equation \cite{Kelton1983, Kelton1991}

\begin{equation}\label{eqn:rateInit}
J = -D e^{-\frac{F(N)}{k_B T}} \frac{\partial}{\partial N} \left( P(N,t) e^{\frac{F(N)}{k_B T}} \right), \tag{$12$}
\end{equation}

where \(J\) is the current or flux across the free energy barrier, in
the cluster-size space.

The solution to the system of equations above yields the familiar CNT
form of the steady-state nucleation rate
\cite{Kelton1983, Kelton1991, Auer2001, Auer2004}

\begin{equation}\label{eqn:rate}
J = \rho_l Z f^+ e^{-\frac{F(N^*)}{k_B T}}, \tag{$13$}
\end{equation}

where the nucleation rate \(J\) is in units of the number of nucleation
events per unit volume per unit time, \(f^+\) is the attachment rate of
molecules to the critical cluster, \(\rho_l\) is the number density of
the supercooled liquid, and \(Z\) is the Zeldovich factor.

The Zeldovich factor is a measure of the curvature at the top of the
free energy barrier. \(Z\) accounts for the fact that postcritical
clusters might still shrink without growing due to re-crossing the
barrier \cite{Pan2004}. According to our formalism, as expressed in the
cluster-size-space, \(Z\) can be calculated using:

\begin{equation}\label{eqn:zeldovich}
Z = \sqrt{\frac{|F''(N^*)|}{2 \pi k_B T}} = \left( \frac{v'}{36\pi N^{*2}} \right)^{\frac{1}{3}} \sqrt{\frac{\sigma_{0} (24G^2+7\eta^2\dot{\gamma}^2)}{6G^2 k_BT}}. \tag{$14$}
\end{equation}

Thus we have derived expressions for the height of the free energy
barrier \(F(N^*)\), and the steady-state nucleation rate \(J\) in terms
of the shear rate, and equilibrium properties \(|\Delta \mu_{0}|\) and
\(\sigma_{0}\) determined in the absence of shear. Later sections show
that these calculations can be performed in a computationally efficient
framework. The other quantities required for the calculation of \(J\)
are \(f^+\), \(v'\), \(\eta\) and \(G\). These input parameters can
either be obtained directly from the seeding technique or estimated
otherwise.

\hypertarget{calculation-of-the-attachment-rate}{%
\subsection{Calculation of the Attachment
Rate}\label{calculation-of-the-attachment-rate}}

According to CNT, the attachment rate \(f^+\) can be related to the time
required for a single particle or molecule to attach to the crystal
cluster \cite{Kelton1983, Kelton1991}. We can estimate \(f^+\) by

\begin{equation}\label{eqn:diffusion1}
f^+ = \frac{6 D_l}{\lambda^2} O_{N^*}, \tag{$15$}
\end{equation}

where \(D_l\) is the diffusion coefficent of the supercooled liquid
phase, \(\lambda\) is the atomic `jump length', estimated to be about
one molecule diameter, and \(O_{N^*}\) is the number of binding sites on
the surface of the critical cluster of size \(N^*\).

\(O_{N^*}\) can be assumed to be equal to the surface area of the
cluster, divided by an area per molecule or particle \cite{Kelton1983}.
The per-particle area in contact with the cluster is assumed to be
\(\pi \left(\frac{3v'}{4 \pi}\right)^{\frac{2}{3}}\) \cite{Kelton1983}.
Hence, we obtain:

\begin{equation}\label{eqn:diffusion2}
O_{N^*} = 4 (N^*)^{\frac{2}{3}} \left( 1 + \frac{7 \eta^2 \dot{\gamma}^2}{24G^2} \right). \tag{$16$}
\end{equation}

Substituting the value of \(O_{N^*}\) from Eq.\eqref{eqn:diffusion2} and
putting it in Eq.\eqref{eqn:diffusion1}, the final expression for
\(f^+\) is

\begin{equation}\label{eqn:diffusionCNT}
f^+ = \frac{24 D_l}{\lambda^2} (N^*)^{\frac{2}{3}} \left( 1 + \frac{7 \eta^2 \dot{\gamma}^2}{24G^2} \right). \tag{$17$}
\end{equation}

We recall that the term \(\frac{7 \eta^2 \dot{\gamma}^2}{24G^2}\) is a
``shape factor'' accounting for the deformation of the cluster due to
the application of shear. For approximately spherical crystallites,
\(\frac{7 \eta^2 \dot{\gamma}^2}{24G^2}\) is negligible. However, the
application of shear still has the overall effect of enhancing \(f^+\)
since both \(D_l\) and \(N^*\) increase with increasing shear rates.

\hypertarget{calculation-of-the-diffusion-coefficient}{%
\subsubsection{Calculation of the Diffusion
Coefficient}\label{calculation-of-the-diffusion-coefficient}}

We expect the diffusion coefficient \(D_l\) to be dependent on the shear
rate \cite{Sandberg1995, Malandro1998, Mura2016, Luo2020}, in addition
to being dependent on the temperature \cite{CusslerDiffusion}. In our
simulations, the applied shear rate is in the \(x\) dimension. Hence, we
can calculate the two-dimensional diffusion coefficient (\(D_l\)) of the
supercooled liquid under the application of shear, in the \(yz\) plane
using the mean square displacement \cite{Li2009}

\begin{equation}\label{eqn:diffusionLiq}
D_l = \lim_{t \to \infty} \frac{1}{4t} \langle [y(t)-y(0)]^2 + [z(t)-z(0)]^2 \rangle, \tag{$18$}
\end{equation}

where \(D_l\) is the two-dimensional self-diffusion coefficient, \(t\)
is the elapsed time with respect to a reference time origin,
\(\langle [y(t)-y(0)]^2 + [z(t)-z(0)]^2 \rangle\) is the ensemble
average over all molecules and time origins, of the mean-squared
displacement of the molecules in the \(y\) and \(z\) dimensions.

\hypertarget{limiting-values-of-shear-rates}{%
\subsection{Limiting Values of Shear
Rates}\label{limiting-values-of-shear-rates}}

The denominator of the free energy in Eq.\eqref{eqn:criticalF} is
undefined for larger shear rates as it tends to shrink to zero. The
correspondingly infinite free energy barrier has well-defined physical
consequences. This situation implies that the clusters are not
mechanically stable at very large shear rates, and this effectively
suppresses nucleation. We define the limiting maximum shear rate
\(\dot{\gamma}_{max}\) as the shear rate for which the free energy
barrier is infinite and for which the nucleation rate effectively
vanishes, given by

\begin{equation}\label{eqn:maxShear}
\dot{\gamma}_{max} = \frac{1}{\eta} \left( \frac{2G |\Delta \mu_{0}|}{v'} \right)^{\frac{1}{2}}, \tag{$19$}
\end{equation}

where \(G\) is the shear modulus in units of pressure,
\(|\Delta \mu_{0}|\) is the difference in chemical potential between the
liquid and solid phases in the absence of shear, \(\eta\) is the
viscosity of the supercooled liquid, and \(v'\) is the volume of a
single molecule in the crystal phase. \(v'\) is, hence, the reciprocal
of the number density of the solid phase.

We note the dependence of \(\dot{\gamma}_{max}\) on the mechanical
properties of the crystal nucleus (\(G\) and \(\eta\)), as well as on
the chemical potential difference and the density of the crystal seed.

Interestingly, \(\dot{\gamma}_{max}\) in Eq.\eqref{eqn:maxShear} is not
dependent on the value of \(N^*\), which implies that for values of
shear rates greater than \(\dot{\gamma}_{max}\) even infinitesmally
small nuclei are mechanically unstable and prone to fragmentation.

Rigorous calculations at every temperature yield the values of
\(|\Delta \mu_0|\) and \(\eta\) at each temperature
\cite{Espinosa2014, Espinosa2015, Dehaoui2015}. However, in order to
determine the dependence of \(\dot{\gamma}_{max}\) on the temperature
\(T\), it is desirable to approximate both \(|\Delta \mu_0|\) and
\(\eta\) by analytical functions of \(T\). Using the enthalpy change of
melting \cite{Kelton1991} and a power law expression for viscosity
\cite{Dehaoui2015}, we can rewrite Eq.\eqref{eqn:maxShear} in terms of
\(T\) by using such analytical approximations for \(|\Delta \mu_0|\) and
\(\eta\), respectively.

The power law expression for \(\eta\) for supercooled water is given by Dehaoui et al.
\cite{Dehaoui2015}

\begin{equation}\label{eqn:powerLawVisco}
\eta = A_0 \left( \frac{T}{T_S} -1 \right)^{-\beta}, \tag{$20$}
\end{equation}

where \(A_0\) and \(T_S\) are fitting parameters with units of dynamic
viscosity and temperature, respectively. \(\beta\) is a dimensionless
fitting parameter.

The difference in the chemical potential, \(\Delta \mu_{0}\), can be
approximated using the enthalpy change at melting \cite{Kelton1991},
which is accurate for low supercoolings and specifically for the
monoatomic (mW) water model \cite{Molinero2009} at larger supercoolings
can be given by:

\begin{equation}\label{eqn:approxMu}
\Delta \mu_{0} = \Delta H_m \left(1 - \frac{T}{T_m} \right), \tag{$21$}
\end{equation}

where \(\Delta H_m\) is the enthalpy change at melting, and \(T_m\) is
the melting point for the model.

Eliminating \(\eta\) and \(|\Delta \mu_0|\) from
Eq.\eqref{eqn:maxShear}, using Eq.\eqref{eqn:powerLawVisco} and
Eq.\eqref{eqn:approxMu}, respectively, we recast \(\dot{\gamma}_{max}\)
as a function of \(T\):

\begin{equation}\label{eqn:expMaxShear}
\dot{\gamma}_{max}(T) = \frac{(T-T_S)^{\beta}}{A_0 T_S^{\beta}} \left( \frac{2G}{v'} \Delta H_m \left( 1-\frac{T}{T_m} \right) \right)^{\frac{1}{2}}. \tag{$22$}
\end{equation}

We can estimate the validity of the approximations made, by comparing
the values of \(\dot{\gamma}_{max}\) calculated using explicitly
determined \(\eta\) and \(|\Delta \mu_{0}|\) values from the literature
\cite{Dehaoui2015, Espinosa2014, Espinosa2015}, with those predicted by
Eq.\eqref{eqn:expMaxShear}.

\begin{figure}[H]
\centering
\includegraphics[scale=0.4]{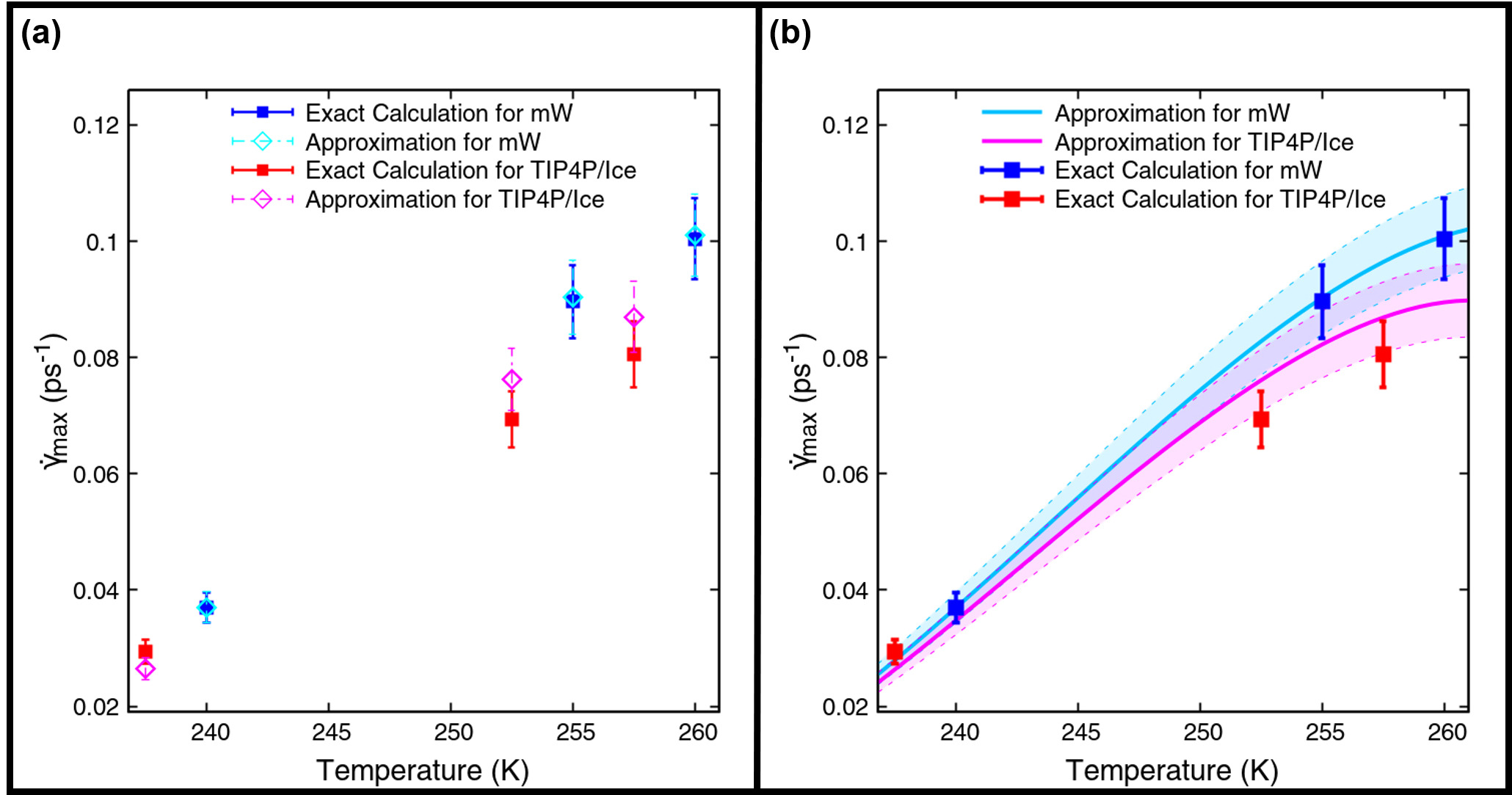}
\caption{(a) Maximum limiting values of the shear rate plotted against temperature, calculated using explicitly determined $\eta$ and $|\Delta \mu_{0}|$ values for the mW (filled blue squares) and TIP4P/Ice (filled red squares) water models \cite{Espinosa2014, Espinosa2015, Dehaoui2015}. Calculations with approximated $\dot{\gamma}_{max}$ estimated from Eq.\eqref{eqn:expMaxShear} for mW and TIP4P/Ice are shown as open diamond symbols in turquioise and magenta, respectively. (b) Comparison of the more exact estimations of $\dot{\gamma}_{max}$ with analytical expressions (Eq.\eqref{eqn:expMaxShear}) using a constant ice density for each water model. The two dotted lines sandwiching the functional approximations for the mW (solid turquoise line) and TIP4P/Ice (solid magenta line) models show the uncertainty ($1$ standard deviation).}
\label{fig:maxShear}
\end{figure}

Figure \ref{fig:maxShear}(a) shows the trend of \(\dot{\gamma}_{max}\)
with temperature for the mW and TIP4P/Ice \cite{Pugliese2017} water
models. Evidently, the the approximate relation in
Eq.\eqref{eqn:expMaxShear} agrees well with the more exact calculations
of \(\dot{\gamma}_{max}\) for the mW model, as expected. Figure
\ref{fig:maxShear}(a) shows that, within the statistical uncertainty,
the values of \(\dot{\gamma}_{max}\) for the mW model nearly coincide.
However, there is more divergence of the approximate values for the
TIP4P/Ice model.

It has been observed that the density of hexagonal ice (Ih) increases
with decreasing temperature at constant pressure
\cite{Vega2009, Vega2011}, which is a trend also reflected in
experiments \cite{Feistel2006}. However, we can further approximate the
value of \(v'\) in Eq.\eqref{eqn:expMaxShear} by assuming that the Ih
density is constant in the temperature range \(235-260 \ K\). Figure
\ref{fig:maxShear}(b) depicts the change in the \(\dot{\gamma}_{max}\)
with temperature, using this additional approximation. The agreement of
the mW model with the predicted approximate values is still excellent.
In the case of the TIP4P/Ice model, the predicted values still differ
from the more exact calculations, but not significantly more than in
Figure \ref{fig:maxShear}(a).

We surmise that Eq.\eqref{eqn:expMaxShear} can be safely used for the
coarse-grained mW water model, but can show small divergences from more
rigorous calculations, for the TIP4P/Ice model. The approximation can be
used to predict general trends and behaviour with tolerable agreement
for TIP4P/Ice.

\hypertarget{methodology-sequence}{%
\subsection{Methodology Sequence}\label{methodology-sequence}}

From Eq.\eqref{eqn:rate}, the quantities required for the estimation of
the nucleation rate \(J\) are the chemical potential difference between
the stable and metastable phases \((|\Delta \mu_{0}|)\), the rate of
attachment of molecules to the critical cluster \((f^+)\), the volume of
a single molecule in the crystal phase at the thermodynamic conditions
\((v')\), the viscosity of the supercooled liquid \((\eta)\) and the
shear modulus of the crystal nucleus \((G)\). We outline the estimation
of the input parameters required for the calculation of \(J\), at
different shear rates:

\begin{enumerate}
\def\labelenumi{\arabic{enumi}.}
\item
  We employ the seeding method to estimate the temperature for known
  critical cluster sizes. Once a trial temperature has been fixed for a
  pre-determined critical cluster size, the interfacial energy
  \(\sigma_0\) can be calculated. In the seeding technique, spherical
  clusters of pure Ih of a certain size are inserted into equilibrated
  configurations of supercooled liquid
  \cite{Espinosa2014, Espinosa2015, Espinosa2016}. For a given inserted
  cluster seed size, the temperature \(T\) for which the seed is
  critical can be determined by running several trajectories and
  tracking the cluster sizes. If the inserted cluster is of critical
  size at the test temperature then the seed grows in roughly half of
  the trajectories, while the cluster shrinks during the course of the
  other runs. The accuracy of this estimation of \(T\) can be improved
  by increasing the number of runs at each trial temperature. Systematic
  errors are larger for smaller cluster sizes owing to the uncertainty
  in the determination of the interfacial region.
\item
  \(|\Delta \mu_{0}|\) can be determined directly by using the Einstein
  molecule method \cite{Vega2007} to compute the chemical potential of
  the fluid and crystal phases \cite{Zaragoza2015}, or by thermodynamic
  integration \cite{Espinosa2014}. Alternatively,
  Eq.\eqref{eqn:approxMu} can be used as an approximation for
  \(|\Delta \mu_{0}|\). As noted previously, Eq.\eqref{eqn:approxMu} is
  accurate only for low supercoolings and differs from rigorously
  calculated \(|\Delta \mu_{0}|\) for atomistic water potentials
  \cite{Espinosa2014}. However, it has been reported that
  Eq.\eqref{eqn:approxMu} performs well for larger supercoolings as well
  \cite{Espinosa2014}, in the case of the mW water model, since the
  maximum of heat capacity is displaced to lower temperatures for this
  model \cite{Molinero2009}.
\item
  \(f^+\) is the rate of attachment of molecules to the critical
  cluster, which can be calculated using Eq.\eqref{eqn:diffusionCNT}. A
  typical attachment or `jump' length \(\lambda\) can be assumed to be
  one molecular diameter \cite{Kelton1991}. For the mW model, we can
  take \(\lambda=3.5\) \si{\angstrom} \cite{Espinosa2014}. The diffusion
  coefficient \(D_l\), which may be dependent on shear, is also required
  for the calculation of \(f^+\).
\item
  \(v'\) is the inverse of the number density of the crystal phase. It
  is known that Ih density varies with the temperature, though the
  extent of this density change varies for different water models
  \cite{Espinosa2014}. The Ih density at every temperature can be
  determined from isothermal-isobaric simulations of the crystal phase.
\item
  \(\eta\) is dependent on temperature, and can either be calculated
  directly from Molecular Dynamics simulations \cite{Hijes2018} or
  estimated by a power law relation \cite{Dehaoui2015}. We note that
  such analytical relations for the viscosity are typically not valid
  for deeply supercooled water below \(230 \ K\). We can assume that
  \(\eta\) is independent of the applied shear rate for `low' shear
  rates; however, at higher shear rates \(\eta\) decreases under shear
  strain. Shear-thinning has been observed for the SPC/E water model for
  shear rates greater than \(0.1 \ ps^{-1}\) \cite{Balasubramanian1996}.
  In this work, the highest limiting shear rate calculated using
  Eq.\eqref{eqn:expMaxShear} is \(\approx 0.1 \ ps^{-1}\), and hence we
  can assume that \(\eta\) is constant with respect to the shear rate,
  at each \(T\), within this limit.
\item
  The shear modulus \(G\) can vary, in the range of
  \(\approx 3-4.5 \ GPa\), for different crystal plane directions and
  cluster structure, for both Ih and amorphous ices
  \cite{Loerting2006, Cao2018, Moreira2019}. Figure S2 shows the
  temperature dependence of \(\dot{\gamma}_{max}\), estimated using
  Eq.\eqref{eqn:expMaxShear} for the mW water model, for different
  values of \(G\). In this work, we have used a constant shear modulus,
  without taking into account the anisotropy of \(G\).
\end{enumerate}

\newpage

\hypertarget{simulation-details}{%
\section{Simulation Details}\label{simulation-details}}

We have applied our formalism to the monatomic (mW) water model
\cite{Molinero2009}, at four temperatures \(235\), \(240\), \(255\) and
\(260 \ K\).

\hypertarget{seeding-method-calculations}{%
\subsection{Seeding Method
Calculations}\label{seeding-method-calculations}}

The seeding method was employed to obtain input parameters, described in
further detail in the Supporting Information. The technique involves the
generation of molecular dynamics trajectories in the absence of shear.
Molecular dynamics simulations of the mW water model were performed
using LAMMPS \cite{Plimpton1995} in the isothermal-isobaric (\(NPT\))
ensemble at \(1 \ atm\). The Nose-Hoover thermostat and barostat were
used to control the temperature and pressure, respectively. A time step
of \(10 \ fs\) was used.

Initial systems were created by inserting a perfectly spherical cluster
of hexagonal ice (Ih) in supercooled liquid configurations obtained from
independent simulations at \(208 \ K\). The liquid molecules which
overlapped with the inserted cluster molecules were removed, and a
tolerance of \(2\) \si{\angstrom} was kept between the cluster and
surrounding supercooled liquid. The systems were then equilibrated in
two steps. First, the solid cluster molecules were kept rigid, allowing
the liquid molecules to move at a temperature of \(200 \ K\) for
\(8 \ ps\). Next, the constraint on the cluster molecules was removed
and the systems were equilibrated for a further \(0.2 \ ns\) at
\(200 \ K\).

The temperature at which each cluster is critical was determined by
running MD trajectories at different temperatures and tracking the
cluster size. The cluster size evolution for a seed size of \(686\)
particles has been shown in Figure S1. The ice-like molecules were
differentiated from the liquid phase by using the structural
identification method of Maras et al. \cite{Maras2016}, in OVITO
\cite{Stukowski2009}. The evolution of the largest ice cluster in each
trajectory was monitored using d-SEAMS \cite{Goswami2020}.

Table S1 lists the input parameters, used in our formalism, for the mW
water model.

\hypertarget{simulations-for-calculating-the-diffusion-coefficient}{%
\subsection{Simulations for Calculating the Diffusion
Coefficient}\label{simulations-for-calculating-the-diffusion-coefficient}}

Molecular dynamics trajectories were obtained at different shear rates
using LAMMPS \cite{Plimpton1995} at \(235\), \(240\), \(255\) and
\(260 \ K\). The two-dimensional diffusion coefficients at different
shear rates were calculated using the VMD Diffusion Coefficient Tool
\cite{Giorgino2019}. Shear was imposed in the \(x\) dimension using the
SLLOD algorithm \cite{Evans1984}, with Lees-Edwards boundary conditions
\cite{Lees1972, Daivis2006}. A timestep of \(5 \ fs\) was used for the
simulations with shear. The system size used for the diffusion
coefficient calculations was \(4096\) molecules of mW water at the four
temperatures considered in this work.

\hypertarget{results-and-discussion}{%
\section{Results and Discussion}\label{results-and-discussion}}

\hypertarget{effect-of-shear-on-diffusion}{%
\subsection{Effect of Shear on
Diffusion}\label{effect-of-shear-on-diffusion}}

We observe that diffusion is enhanced by the applied shear, as expected
from previous results in the literature
\cite{Sandberg1995, Malandro1998, Mura2016, Luo2020}. By fitting the
two-dimensional bulk diffusion coefficients at different shear rates, we
obtain the following linear relationship at a constant temperature
\(T\):

\begin{equation}\label{eqn:expDiffusion}
D_l = D_0 + c \dot{\gamma}, \tag{$23$}
\end{equation}

where \(D_l\) is the two-dimensional diffusion coefficient at a
particular shear rate \(\dot{\gamma}\) and temperature \(T\), \(D_0\) is
the bulk diffusion coefficient in the absence of shear at \(T\), and
\(c\) is a fitting constant corresponding to the rate of increase of the
diffusion coefficient with shear, which has units of squared length.

Figure S3 shows the linear fits of the diffusion coefficient with shear
rates, for the four temperatures studied in this work.

\begin{figure}[H]
\centering
\includegraphics[scale=0.4]{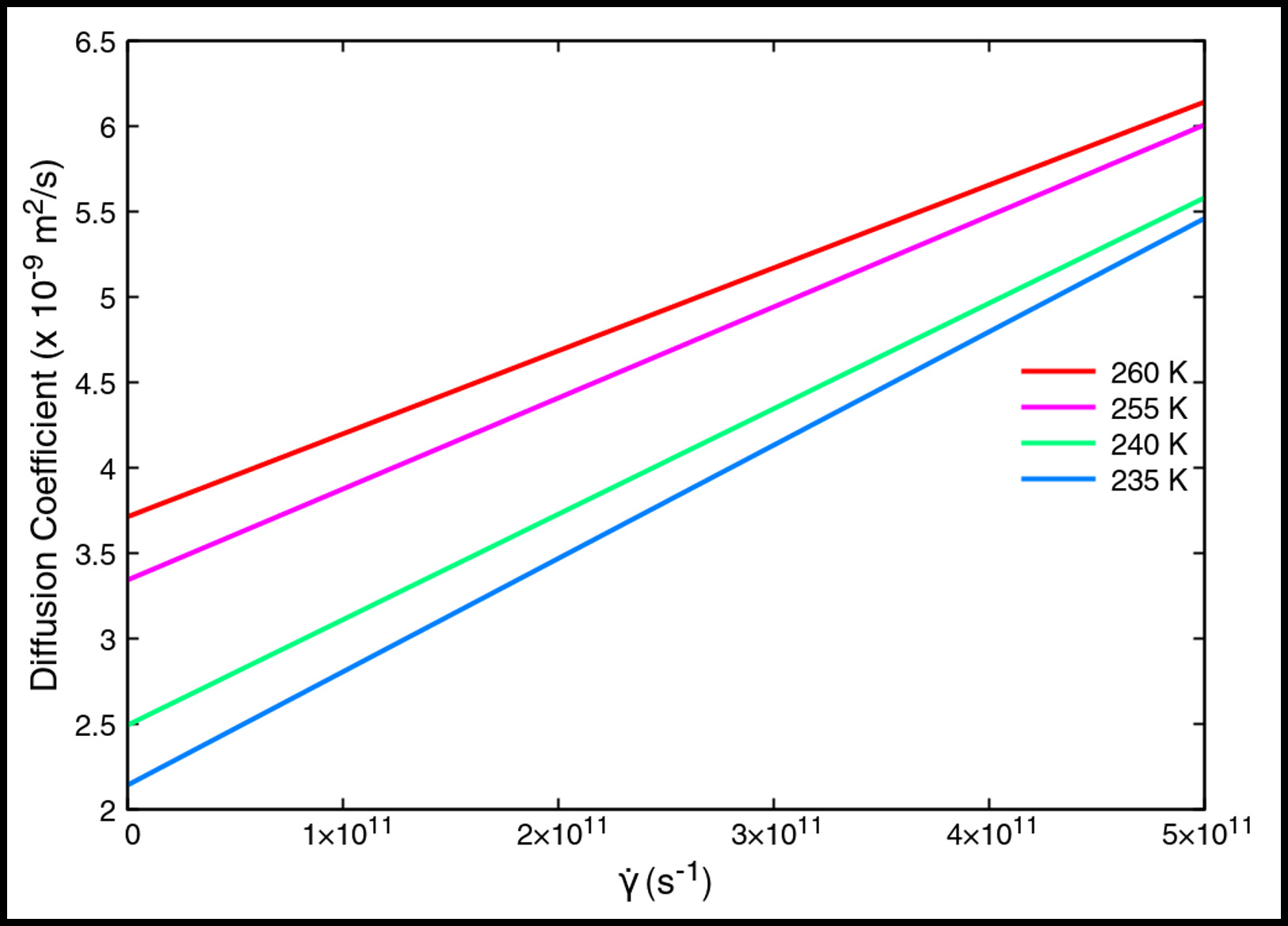}
\caption{Variation of the Diffusion coefficient with shear, for four different temperatures. It is observed that the Diffusion coefficients increase linearly with shear.}
\label{fig:diffusion}
\end{figure}

Figure \ref{fig:diffusion} depicts the fitted lines for each
temperature. We observe that, in general, the diffusion coefficient
increases with increasing temperature, as expected. We also surmise that
the rate of increase of the diffusion coefficient \(c\) decreases with
increasing temperature. Table \ref{params} lists the values of \(D_0\)
and \(m\) for each temperature.

\begin{longtable}[]{@{}lll@{}}
\caption{Parameter values in \(D_l = D_0 + c\dot{\gamma}\)
\label{params}}\tabularnewline
\toprule
Temperature & \(D_0\) (\(\times 10^{-9} \ m^2/s\)) & c
(\(\times 10^{-21} \ m^2\))\tabularnewline
\midrule
\endfirsthead
\toprule
Temperature & \(D_0\) (\(\times 10^{-9} \ m^2/s\)) & c
(\(\times 10^{-21} \ m^2\))\tabularnewline
\midrule
\endhead
260 K & 3.715 & 4.85364\tabularnewline
255 K & 3.345 & 5.32611\tabularnewline
240 K & 2.496 & 6.17091\tabularnewline
235 K & 2.145 & 6.63036\tabularnewline
\bottomrule
\end{longtable}

\hypertarget{variation-of-the-nucleation-rate-with-shear-rate}{%
\subsection{Variation of the Nucleation Rate with Shear
Rate}\label{variation-of-the-nucleation-rate-with-shear-rate}}

The application of shear has two opposing contributions to the
nucleation rate: the free energy barrier increases with the shear rate,
tending to retard nucleation, while the kinetic pre-factor rises with
shear. The increase in the kinetic pre-factor tends to promote
nucleation. We also note that the the critical nucleus size also
increases with the shear rate \cite{Mokshin2013, Mura2016}, shown in
Figure S4.

\begin{figure}[H]
\centering
\includegraphics[scale=0.45]{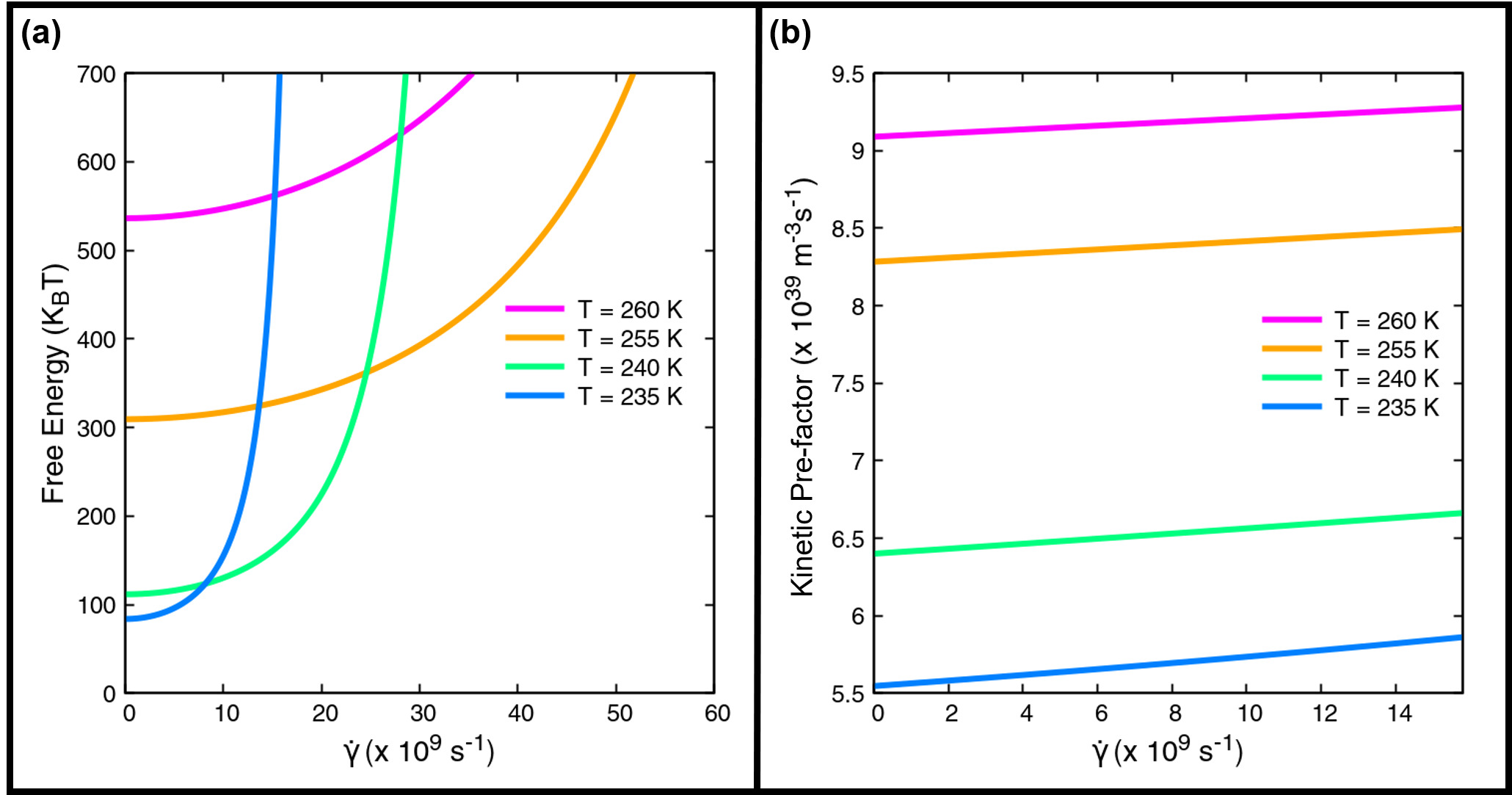}
\caption{(a) Variation of the free energy barrier $F(N^*)$ with the shear rate $\dot{\gamma}$ for $260$, $255$, $240$ and $235$ K. (b) The kinetic pre-factor $\rho_l Z f^+$ plotted against the shear rate $\dot{\gamma}$ for different temperatures. }
\label{fig:kineticVsExpo}
\end{figure}

Figure \ref{fig:kineticVsExpo}(a) shows the quadratic increase of the
free energy barrier with rising shear rates. The quadratic increase in
the free energy barrier of nucleation with shear has also been reported
for colloidal suspensions \cite{Blaak2004}. Figure
\ref{fig:kineticVsExpo}(b) depicts the increase of the kinetic
pre-factor with shear for different temperatures. The diffusion
coefficients were calculated using Eq. \ref{eqn:expDiffusion}. We
surmise that the non-monotonicity of the nucleation rate arises from the
interplay of these two conflicting tendencies.

\begin{figure}[H]
\centering
\includegraphics[width=\textwidth]{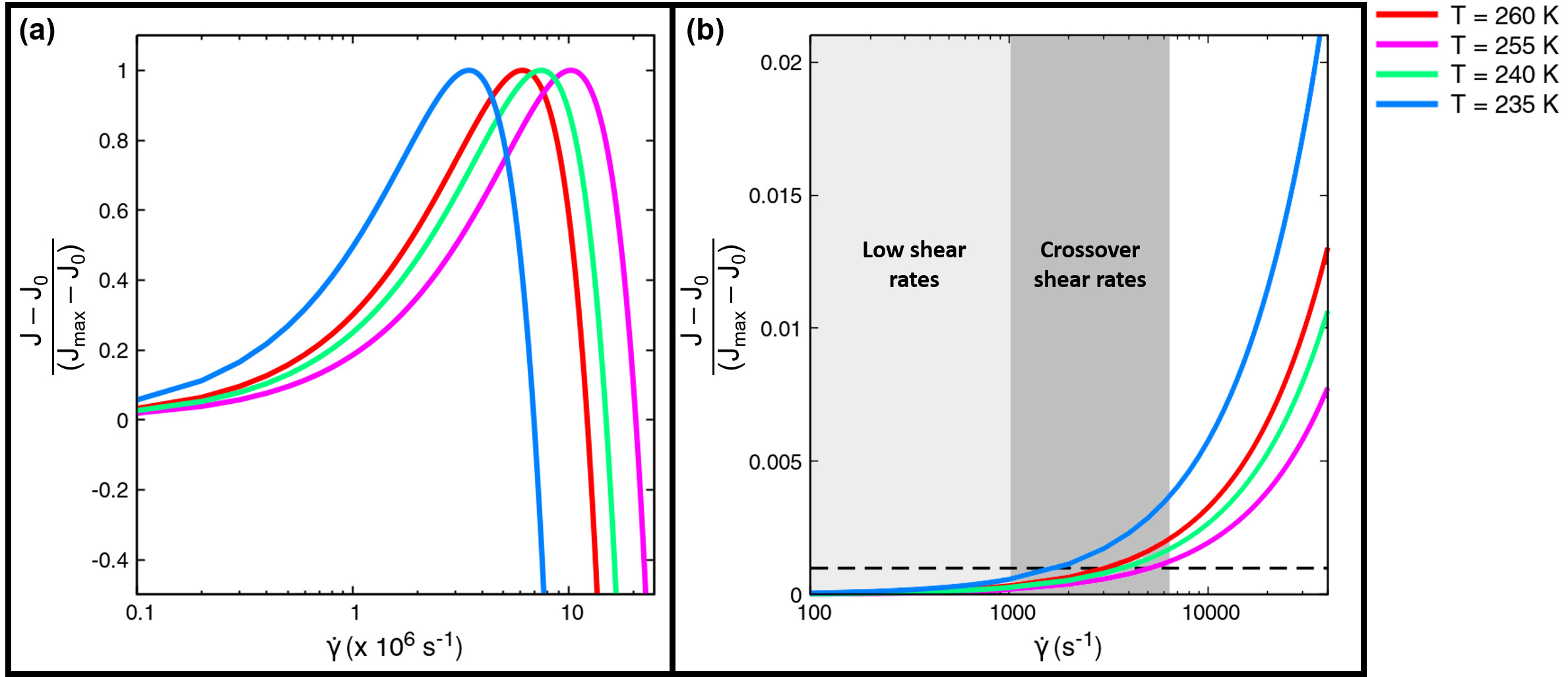}
\caption{(a) Variation of the normalized relative nucleation rates $\frac{J-J_0}{J_{max}-J_0}$ with shear, for shear rates greater than $10^5 \ s^{-1}$, for different temperatures. The non-monotonic behaviour arises from the competing energetic and kinetic effects of shear. (b) Normalized relative nucleation rates, plotted against the shear rate, depicting the regimes of low shear rates and crossover shear rates. In the low shear rate regime, upto $1000 \ s^{-1}$, highlighted in light grey, the nucleation rates are unaffected by the applied shear. The nucleation rates begin to increase for shear rates greater than the crossover shear rates, shaded in dark grey. The black dotted line denotes the normalized relative nucleation rate value of $0.001$, for which the crossover shear rates are defined.}
\label{fig:nucleationRates}
\end{figure}

Figure \ref{fig:nucleationRates}(a) depicts the non-monotonicity of the
dimensionless relative nucleation rates \(\frac{J-J_0}{J_{max}-J_0}\)
with shear, for four different temperatures. This dimensionless relative
nucleation rate is defined with respect to \(J_0\), the nucleation rate
calculated when there is no shear, and \(J_{max}\). \(J_{max}\) is the
highest nucleation rate, maximized with respect to the shear rate, at a
particular temperature. Figure S5 shows \(J_{max}\) obtained for the
four temperatures considered in this work. Figure
\ref{fig:nucleationRates}(b) shows plots of the normalized relative
nucleation rate, in the limit of low shear rates.

Our results indicate the existence of three qualitative regimes of
crystallization behaviour under shear. At low shear rates, in the range
of \(0-1000 \ s^{-1}\), there is negligible effect on the kinetics
(highlighted in light grey in Figure \ref{fig:nucleationRates}(b)). Such
shear rates are typically imposed in Couette flows in experiments
\cite{2009a}.

However, for shear rates between \(1000-10,000 \ s^{-1}\), the
nucleation rates begin to increase smoothly.  We define a crossover shear rate, indicative of this smooth transition, such that at the crossover shear rate the normalized relative nucleation rate is $0.001$. The region of crossover shear rates for different
temperatures is highlighted in dark grey in Figure
\ref{fig:nucleationRates}(b).

The nucleation rate continues to rise steadily for higher shear rates
for every temperature, upto shear rates of the order of
\(10^6 \ s^{-1}\). Between \(\approx 10^6-10^7 \ s^{-1}\), a maximum in
the nucleation rate is observed, originating from the competing kinetic
and energetic effects of shear (Figure \ref{fig:nucleationRates}(a)). We
define the shear rate at which the nucleation rate is maximum as the
optimal shear rate, at a particular temperature. Beyond the optimal
shear rate, very high shear rates inhibit nucleation, until the
nucleation rate vanishes. In the regime of very high shear rates larger
than the optimal shear rate, the free energy cost of nucleation
dominates the enhancement in diffusion due to the shear.

We also note that the optimal shear rates observed are much higher than
shear rates typically used in experiments. The range of shear rates
generated in experiments can be about \(0-100,000 \ s^{-1}\)
\cite{Kumar2016}, orders of magnitude lower than the optimal shear
rates.

\hypertarget{dependence-of-the-optimal-shear-rates-and-nucleation-rate-curves-on-temperature}{%
\subsection{Dependence of the Optimal Shear Rates and Nucleation Rate
Curves on
Temperature}\label{dependence-of-the-optimal-shear-rates-and-nucleation-rate-curves-on-temperature}}

The behaviour of the nucleation rate curves in Figure
\ref{fig:nucleationRates} indicates a possible non-linear dependence on
the temperature. This is due to the inclusion of several
temperature-dependent parameters in the expression for the nucleation
rate (Eq.\eqref{eqn:rate}). An analytical expression for the optimal
shear rate cannot be obtained because of the dependence of the
exponential term and kinetic pre-factor on the shear rate.

\begin{figure}[H]
\centering
\includegraphics[scale=0.45]{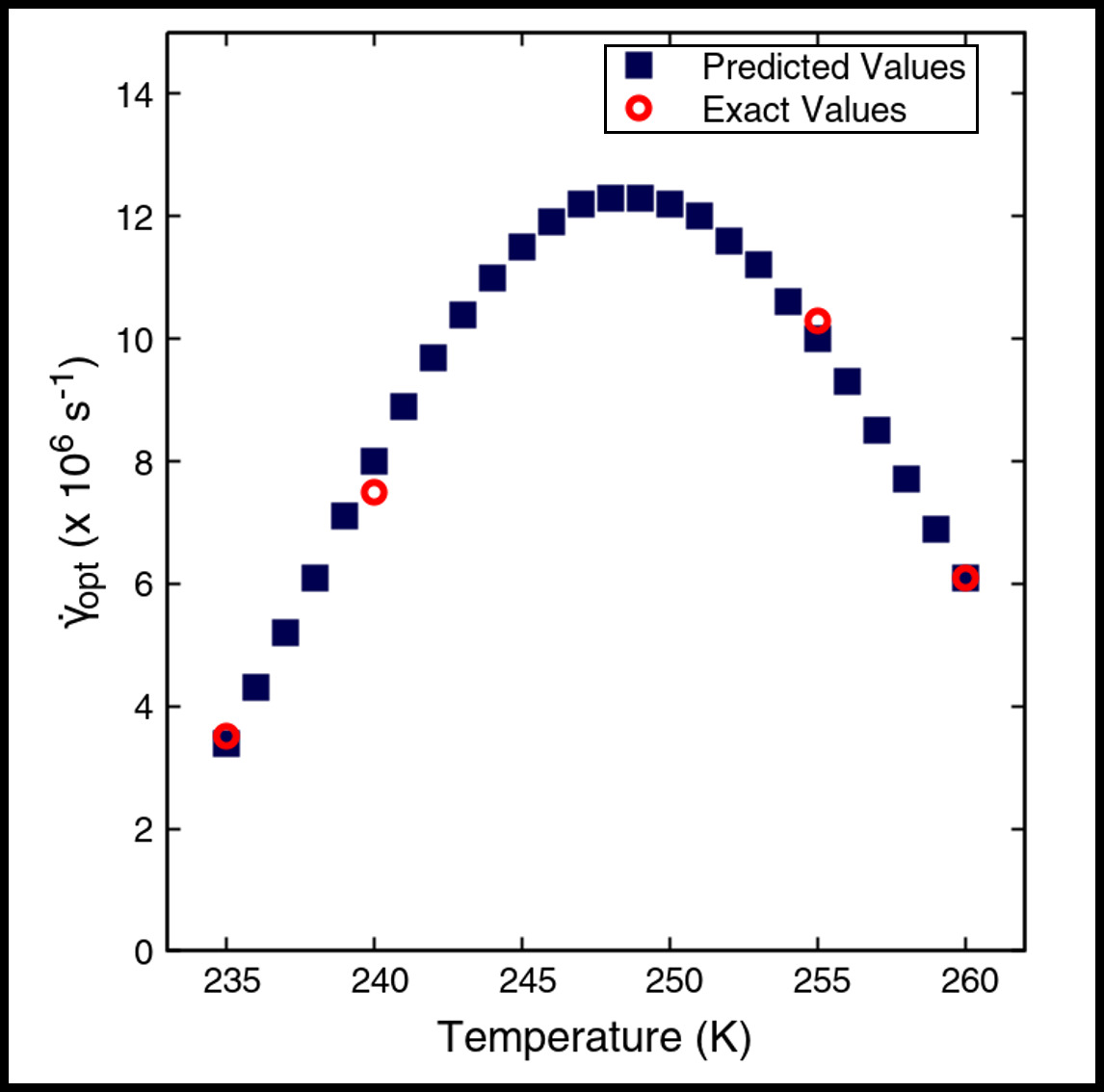}
\caption{Temperature dependence of the predicted and exact values of the optimal shear rates $\dot{\gamma}_{opt}$. The predicted optimal shear rates and the optimal shear rates calculated from seeded simulations are denoted by filled cobalt blue squares and open red circles, respectively. The maximum value of the optimal shear rate is observed at $248 \ K$.}
\label{fig:maxPredictOptShear}
\end{figure}

In order to validate the functional form of the trend from our four
simulations, we employ statistical inference. We qualitatively ascertain
the non-linear temperature dependence of the nucleation rates and
optimal shear rates, by extrapolating data for different temperatures
using the approximations outlined below. The validity of the predicted
results can be assessed by comparing them with the more rigrously
obtained data described in the previous section.

\begin{enumerate}
\def\labelenumi{\arabic{enumi}.}
\tightlist
\item
  The interfacial energy \(\sigma_0\) for all temperatures is assumed to
  be constant and taken to be the mean of the values calculated from
  seeding in Table S1. We have used this assumption with the
  understanding that the maximum variation in the calculated values of
  \(\sigma_0\) is within \(1.5 \%\), and that the statistical
  uncertainty in the interfacial energy calculations is \(\approx 7 \%\)
  \cite{Espinosa2014}.
\item
  The diffusion coefficients are assumed to vary with shear according to
  a linear relationship given by Eq.\eqref{eqn:expDiffusion}, at every
  temperature. The values of the parameters \(D_0\) and \(c\), obtained
  in Section 4.1, are estimated for each temperature from linear
  regression. The linear fits of \(D_0\) and \(c\) with temperature are
  shown in Figure S6.
\item
  The chemical potential \(|\Delta \mu_0|\) and the viscosity \(\eta\)
  at each temperature are approximated by Eq.\eqref{eqn:approxMu} and
  Eq.\eqref{eqn:powerLawVisco}, respectively.
\item
  The density of Ih is assumed to be constant in the range of
  temperatures \(235 \ K\) to \(260 \ K\).
\end{enumerate}

Using the approximations detailed above, we obtained the optimal shear
rates for incrementally increased temperatures between \(235 \ K\) and
\(260 \ K\). Figure \ref{fig:maxPredictOptShear} shows the variation of
the predicted optimal shear rates with temperature. The more exact
calculations for \(235\), \(240\), \(255\) and \(260 \ K\) are plotted
alongside the predicted values, showing agreement within \(10 \%\). It
is evident that the predicted optimal shear rates show a non-monotonic
dependence on the temperature, exhibiting a maximum at \(248 \ K\). The
values of both the predicted and exact optimal shear rates at
\(260 \ K\) are less than those at \(255 \ K\) and \(240 \ K\). The
agreement of the predicted nucleation rate curves and the nucleation
rate curves obtained in the previous section from seeding calculations
is graphically shown in Figure S7.

\begin{figure}[H]
\centering
\includegraphics[width=\textwidth]{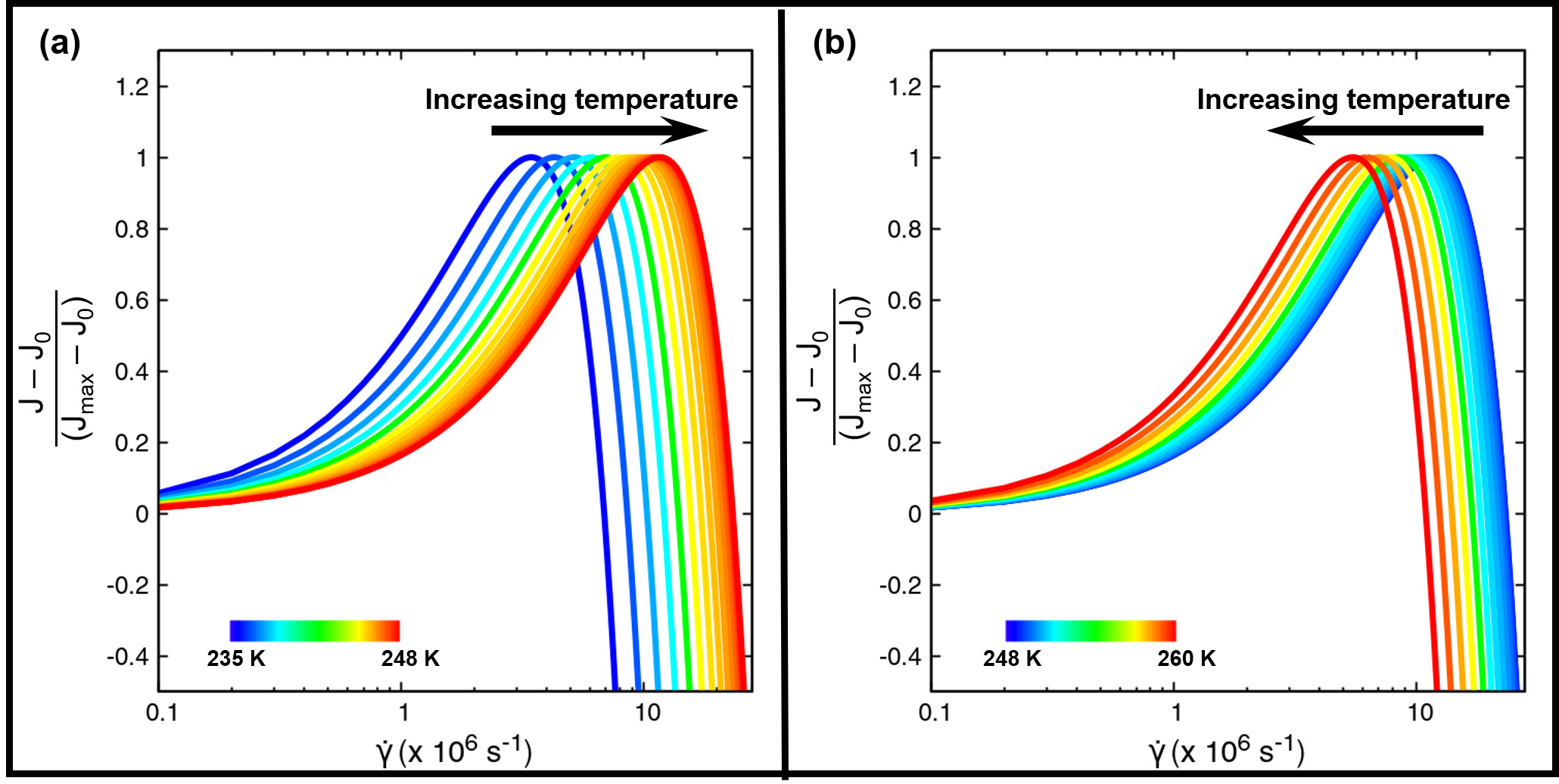}
\caption{(a) Predicted normalized relative nucleation rate curves, varying with the shear rate $\dot{\gamma}$, for temperatures between $235 \ K$ and $248 \ K$. The optimal shear rate at each temperature increases from $235 \ K$ and reaches a maximum value at $248 \ K$. (b) Shear-dependent predicted nucleation rate curves for $248 \ K$ to $260 \ K$. Arrows signify the direction of increasing temperature.}
\label{fig:predictNucRates}
\end{figure}

Figure \ref{fig:predictNucRates}(a) depicts the predicted nucleation
rate curves, plotted against the shear rate, for temperatures between
\(235 \ K\) and \(248 \ K\). The nucleation rate curves are shifted in
the \(+x\) direction with increasing temperature. The nucleation rate
curve at \(248 \ K\) has the highest optimal shear rate and is the
rightmost curve. Figure \ref{fig:predictNucRates}(b) shows the
nucleation rate curves from \(248 \ K\) to \(260 \ K\). As the
temperature increases from \(248 \ K\), the curves are shifted in the
\(-x\) direction. Thus, the predicted results reinforce the suggestion
of the non-monotonic temperature dependence of both the nucleation rates
and the optimal shear rates.

However, although the approximations and predicted data offer insight
into the non-linear temperature dependence, there are a few unavoidable
caveats. Particularly the approximation in Eq.\eqref{eqn:approxMu}, used
for the chemical potential at different temperatures, is more-or-less
valid over a wide range of temperatures for the mW water model only.
Presumably, this approximation would only be acceptable for atomistic
water potentials in the limit of strong supercooling. Unentangling the
complex composite effects of temperature on the nucleation rate is
non-trivial using more rigorous calculations, and requires further
investigation.

\hypertarget{dependence-of-the-maximum-shear-rate-on-temperature}{%
\subsection{Dependence of the Maximum Shear Rate on
Temperature}\label{dependence-of-the-maximum-shear-rate-on-temperature}}

We recall that, at a particular temperature \(T\), nucleation is
suppressed for shear rates greater than the maximum limiting shear rate
\(\dot{\gamma}_{max}\). We can use the expression given by
Eq.\eqref{eqn:expMaxShear} to obtain the maximum shear rate
\(\dot{\gamma}_{max}\) as a function of the temperature \(T\).

\begin{figure}[H]
\centering
\includegraphics[scale=0.45]{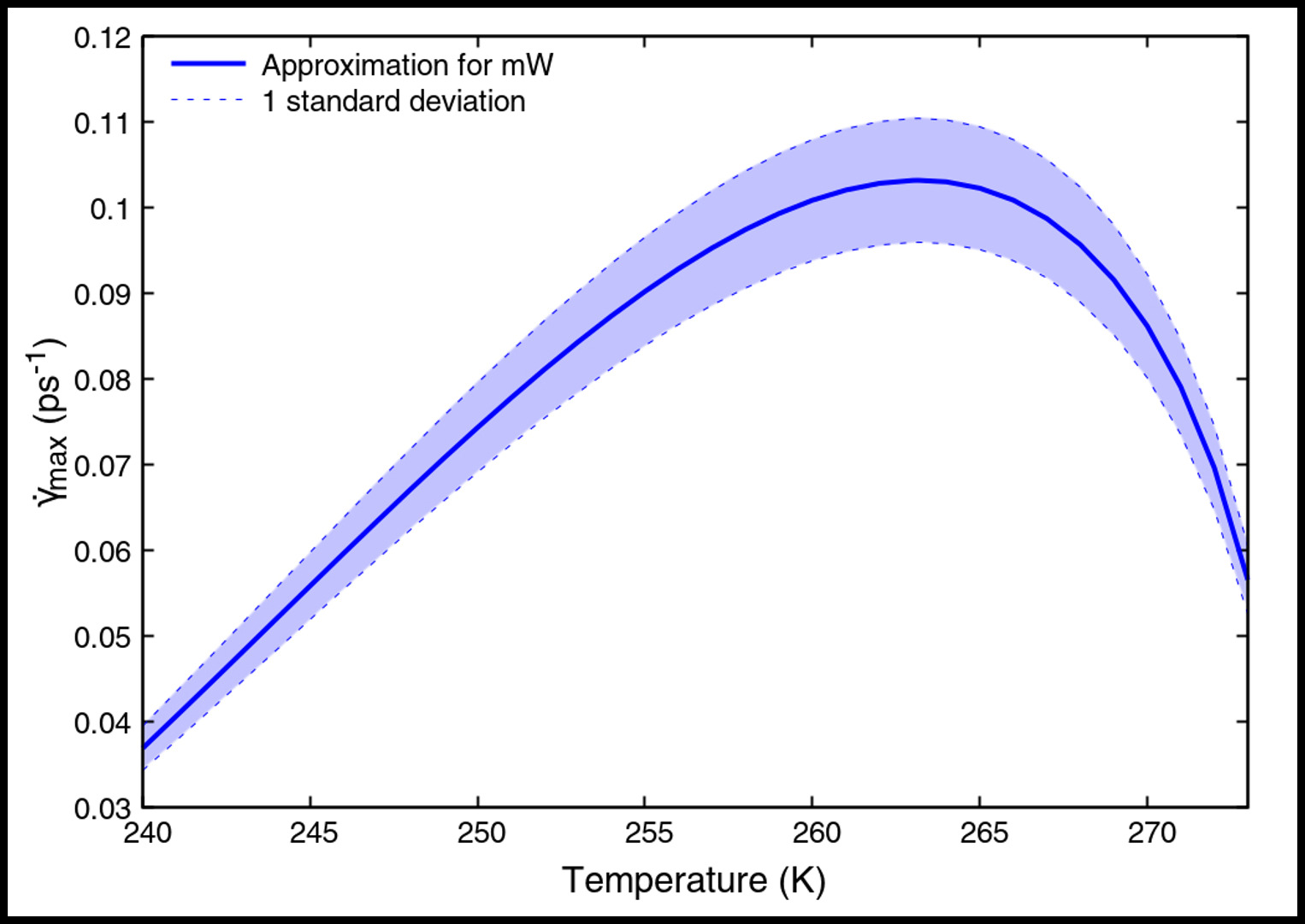}
\caption{Maximum limiting values of the shear rate plotted against temperature for the mW water model, estimated using Eq.\eqref{eqn:expMaxShear}. The dotted blue lines denote the uncertainty in the calculation of $\dot{\gamma}_{max}$.}
\label{fig:maxShearCritical}
\end{figure}

Figure \ref{fig:maxShearCritical} shows the variation of
\(\dot{\gamma}_{max}\) with \(T\) for the mW water model. Eq.
\eqref{eqn:expMaxShear} is a smooth, continuous and differentiable
function in \(T\), and exhibits a maximum for a critical value of
temperature, as shown in Figure \ref{fig:maxShearCritical}.

Maximizing Eq.\eqref{eqn:expMaxShear} with respect to \(T\), we obtain a
critical \(T_c\) for which the \(\dot{\gamma}_{max}\) is maximum, given
by

\begin{equation}\label{eqn:expMaxShearMaxima}
T_c = \frac{2\beta}{2\beta+1}(T_m+T_S), \tag{$24$}
\end{equation}

where \(\beta\) and \(T_S\) are fitting parameters from
Eq.\eqref{eqn:powerLawVisco}, and \(T_m\) is the melting point.

\(T_c\) is the temperature at which crystal nuclei survive at the
highest shear rate, for a particular water model.

Thus, the maximum shear rate, \(\dot{\gamma}_{c,max}\), beyond which
nucleation is suppressed at all temperatures is given by the maxima of
Eq.\eqref{eqn:expMaxShear}, and we obtain:

\begin{equation}\label{eqn:biggestShearRateModel}
\dot{\gamma}_{c,max} = \frac{1}{A_0} \left( \frac{2\beta}{T_S} \right)^{\beta} \left( \frac{T_m-T_S}{2\beta+1} \right)^{\beta+\frac{1}{2}} \sqrt{\frac{2G\Delta H_m}{v' T_m}}, \tag{$25$}
\end{equation}

where \(A_0\) is a fitting parameter in Eq.\eqref{eqn:powerLawVisco}
with the units of dynamic viscosity, and \(\Delta H_m\) is the enthalpy
change of melting.

We note that there is no dependence of \(T_c\) on the shear modulus
\(G\). This is reaffirmed by the plots of \(\dot{\gamma}_{max}\) with
\(T\) in Figure S2, wherein \(T_c = 263.2 \ K\) for all three shear
moduli.

\hypertarget{conclusions}{%
\section{Conclusions}\label{conclusions}}

In this work, we have derived an extension to the CNT equations,
explicitly accounting for shear. We have presented a biasing technique
for studying the effect of shear on nucleation rates, combined with the
seeding method, for studying nucleation at moderate supercooling. At
such temperatures, brute-force MD simulations can be infeasible,
especially since the shear rate increases the barrier height.

Our results reveal the existence of different shear regimes of
nucleation behaviour for the mW water model: 1) low shear rates in the
range of \(0-1000 \ s^{-1}\), for which the nucleation rate is
unaffected; 2) high shear rates in the range of \(10^6-10^7 \ s^{-1}\),
in which range a maxima in the nucleation rate is observed; 3) very high
shear rates for which nucleation is retarded. We identify crossover
shear rates indicating the increase of the nucleation rate with shear in
the range of \(1000-10,000 \ s^{-1}\). We also define optimal shear
rates as those corresponding to the maximum in the nucleation rate at a
particular temperature. Our formalism is thus capable of predicting
trends over a wide range of shear rates, in a relatively computationally
inexpensive manner. Experiments support that there is little effect on
nucleation in the regime of low shear rates
\cite{Hansen2001, Akio1992, 2017a}. The non-monotonicity and decrease in
nucleation rates due to shear agree with previous simulation studies in
the literature
\cite{Blaak2004, Mokshin2013, Richard2015, Mura2016, Luo2020}. The
non-monotonicity arises from the competing kinetic and energetic effects
of shear on the nucleation rate.

The behaviour of the shear-dependent nucleation rates at each
temperature hints at a possible non-linear dependence on temperature.
Using various approximations, we predict that the optimal shear rates
and nucleation rates have a non-monotonic dependence on temperature. We
show that the predicted data agree with the results obtained from seeded
simulations reasonably well enough to make definite assessments. In the
future, further investigations could reveal the role of temperature in
the complex interplay of factors influencing the nucleation rate.

We have calculated the analytical maximum limiting shear rate, beyond
which the nucleation rate vanishes. At the maximum limiting shear rate,
nucleation is completely suppressed due to the mechanical failure of
incipient crystallites. We observe that the maximum limiting shear rate
also has a non-monotonic dependence on temperature, similar to the
optimal shear rates.

Although we have studied the mW water model in detail in this work, the
formalism presented is general and applicable to other systems as well.
By using analytical approximations within the framework of CNT, we are
able to qualitatively describe the nucleation behaviour of simply
sheared systems without resorting to computationally expensive
simulations. Our results indicate a tantalizing, hitherto unexplored,
non-monotonic temperature dependence of the shear-dependent nucleation
rate curves of water. We envisage that this formalism can be used to
provide insight into the qualitative nucleation behaviour of flowing
systems inaccessible to brute-force MD and equilibrium rare-event
acceleration techniques.

\hypertarget{acknowledgements}{%
\section{Acknowledgements}\label{acknowledgements}}

This work was supported by the Science and Engineering Research Board
(sanction number STR/2019/000090 and CRG/2019/001325). Computational
resources were provided by the HPC cluster of the Computer Center (CC),
Indian Institute of Technology Kanpur.

\printbibliography


\end{document}